\documentclass[a4paper,20pt]{article}
    \usepackage[T1]{fontenc}
    \usepackage{graphicx}
    \usepackage{epsfig}
    \usepackage{amsmath}
    \usepackage{amsfonts}
    \renewcommand{\abstract}{}
\begin{document}
\makeatletter
\renewcommand{\@oddhead}{\textit{YSC'14 Proceedings of Contributed Papers} \hfil \textit{V.N. Melnik, A.A. Konovalenko, et al.}}
\renewcommand{\@evenfoot}{\hfil \thepage \hfil}
\renewcommand{\@oddfoot}{\hfil \thepage \hfil}
\fontsize{11}{11} \selectfont

\title{Decameter Type III-Like Bursts}
\author{\textsl{V.N. Melnik$^{1}$, A.A. Konovalenko$^{1}$,
B.P. Rutkevych$^{1,}$}\\
\textsl{H.O. Rucker$^{2}$, V.V. Dorovskyy$^{1}$, E.P. Abranin$^{1}$,}\\
\textsl{A. Lecacheux$^{3}$, A.I. Brazhenko$^{4}$, A.A.Stanislavskyy$^{1}$}}
\date{}
\maketitle
\begin{center} {\small
$^{1}$Institute of Radio Astronomy, Ukrainian Academy of Sciences, Kharkov, Ukraine,  \\
$^{2}$Space Research Institute, Austrian Academy of Sciences, Graz, Austria,\\
$^{3}$Departement de Radioastronomie, Observatoire de Paris, Paris, France,  \\
$^{4}$Poltava Gravimetrical Observatory, Poltava, Ukraine.\\
bprutkevich@mail.ru }
\end{center}

\begin{abstract}
Starting from 1960s Type III-like bursts (Type III bursts with high
drift rates) in a wide frequency range from 300 to 950MHz have been
observed. These new bursts observed at certain frequency being
compared to the usual Type III bursts at the same frequency show
similar behaviour but feature frequency drift 2-6 times higher than
the normal bursts. In this paper we report the first observations of
Type III-like bursts in decameter range, carried out during summer
campaigns 2002 - 2004 at UTR-2 radio telescope. The circular
polarization of the bursts was measured by the radio telescope
URAN-2 in 2004. The observed bursts are analyzed and compared with
usual Type III bursts in the decameter range. From the analysis of
over 1100 Type III-like bursts, their main parameters have been
found. Characteristic feature of the observed bursts is similar to
Type III-like bursts at other frequencies, i.e. measured drift rates
(5-10~MHz/s) of this bursts are few times larger than that for usual
Type III bursts, and their durations (1-2~s) are few times smaller
than that for usual Type III bursts in this frequency band.
\end{abstract}

\section*{Introduction}
\indent \indent For the first time Type III-like bursts were
observed in the frequency range 500 - 950~MHz \cite{Young61}. Their
frequency rates were higher than 2000~MHz/s and sometimes even
infinite. The author also reported bursts with reverse drifts (from
low to high frequencies). The observed number of Type III-like
bursts was 20\% of normal Type III bursts. Durations of fast bursts
ranged from 0.3~s to 2~s. According to \cite{Young61}, their radio
fluxes had values from 5~s.f.u. to 50000~s.f.u. There was a tendency
that the faster Type III-like bursts had shorter duration.

In paper \cite{Kundu61} it was found that Type III-like bursts appeared
mainly in the frequency band 400-800~MHz, but sometimes this band was as
wide as 200-950~MHz. There was information \cite{Gopala_Rao65} that Type
III-like bursts were observed at frequencies 40~MHz and 60~MHz.

Detailed analysis of properties of Type III-like bursts observed at
frequencies 310-340 MHz was performed in \cite{ Elgaroy80,
Elgaroy74}. It was found that 50\%, 17\% and 33\% of Type III-like
bursts out of 402 analyzed ones had forward, reverse and infinite
drift rates, respectively. The mean duration of Type III-like bursts
was 0.26 s which is 1/4 of that for usual meter range Type III
bursts. Paper \cite{Elgaroy80} showed that frequency drift rates and
durations of these bursts depended on the position of the active
region associated with these bursts on the solar disk. From this
fact the author concluded that propagation effects play a
significant role in generation of these bursts.

An attempt to find Type III-like bursts in decameter range was reported in
paper \cite{Zaitsev84}, where the authors analyzed the storm observed at
UTR-2 radio telescope on June, 8-9, 1977. Authors concluded that if such
bursts existed they had drift rates and durations close to that for usual
Type III bursts.

In this paper we present results of observations of decameter (10-30~MHz)
Type III-like bursts at radio telescope UTR-2 performed in 2002-2004 and
URAN-2 in 2004, we also analyze their parameters and discuss their properties.

\section*{Observations}
\indent \indent Type III-like bursts reported in this paper were
registered at radio telescope UTR-2 during summer campaigns of
2002-2004 years. Three sections of the radio telescope, with
effective area of 30000 m$^{2}$ providing the beam of $1^{ \circ}
\times 13^{ \circ} $ were used. In 2002 the registrations were
carried out by a Digital Spectrum Polarimeter (DSP) with frequency
resolution 12 kHz, time resolution 20~ms, 50~ms and 100~ms, and
sensitivity 5 Jy. The frequency band in these observations was 12
MHz. The measurements in 2003-2004 were carried out by a 60-channel
spectrometer with frequency resolution 3 kHz, and time resolution of
10 ms at frequency range 10-30 MHz.

According to the observations in the decameter range (10 - 30 MHz), the
usual Type III bursts have frequency drift rates of 2-4~MHz/s and durations
of 4 - 10 s. Besides these bursts we registered fast Type III bursts, with
drift rates exceeded, sometimes considerably, those for typical Type III
bursts. Following \cite{Elgaroy80}, these fast bursts can be called as Type
III-like bursts. In this report we consider the bursts observed in 2002
(July 13-17, July 26-30 and August 16-19), 2003 (July 1-6), and in 2004
(June 17-22).

\begin{figure}
\begin{center}
\includegraphics[width=12cm]{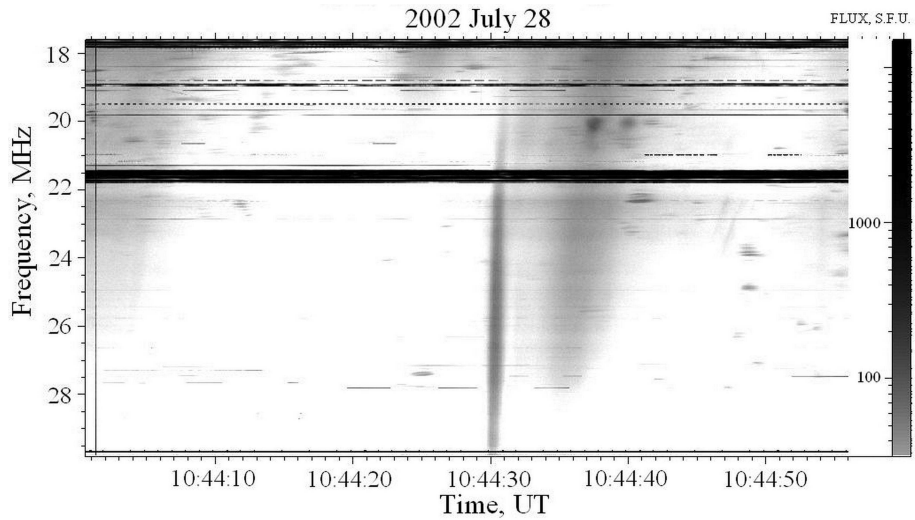}
\caption{ Dynamic spectrum of Type III-like burst. Three bright
vertical areas at 10:43:40, 10:44:35 and 10:45:00 correspond to
normal Type III bursts, and a narrow bright line at 10:44:30
indicates a Type III-like burst. \label{Fig1}}
\end{center}
\end{figure}

\begin{figure}
\begin{center}
\includegraphics[width=12cm]{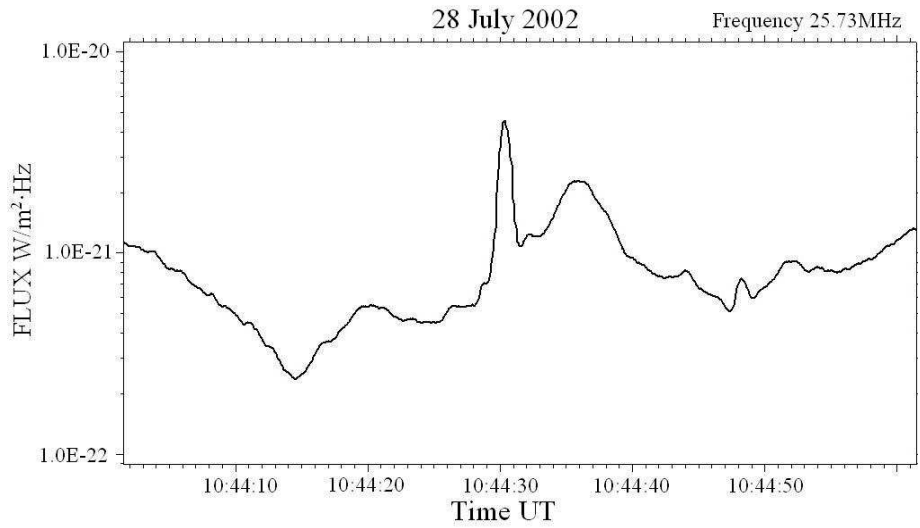}
\caption{ Time profiles of Type III-like burst at the frequency 22.11 MHz. Noticeable symmetrical shape.
\label{Fig1b}}
\end{center}
\end{figure}

The total time of observations was approximately 120 hours and there were
detected about 1100 Type III-like bursts. All bursts had negative frequency
drift rates (from high to low frequencies). Decameter Type III-like bursts
similarly to their high frequency analogs were shorter than usual Type III
bursts (Figure~\ref{Fig1}), and their time profiles were symmetrical
(Figure~\ref{Fig1b}). There were fast bursts with fine frequency structure
similar to Type IIIb bursts, which we call Type IIIb-like bursts. The
fastest Type III-like burst was observed on August 18, 2002 with the drift
rate of about 40~MHz/s.

\begin{figure}
     \begin{center}
     $
     \begin{array}{cc}
     \includegraphics[width=8cm]{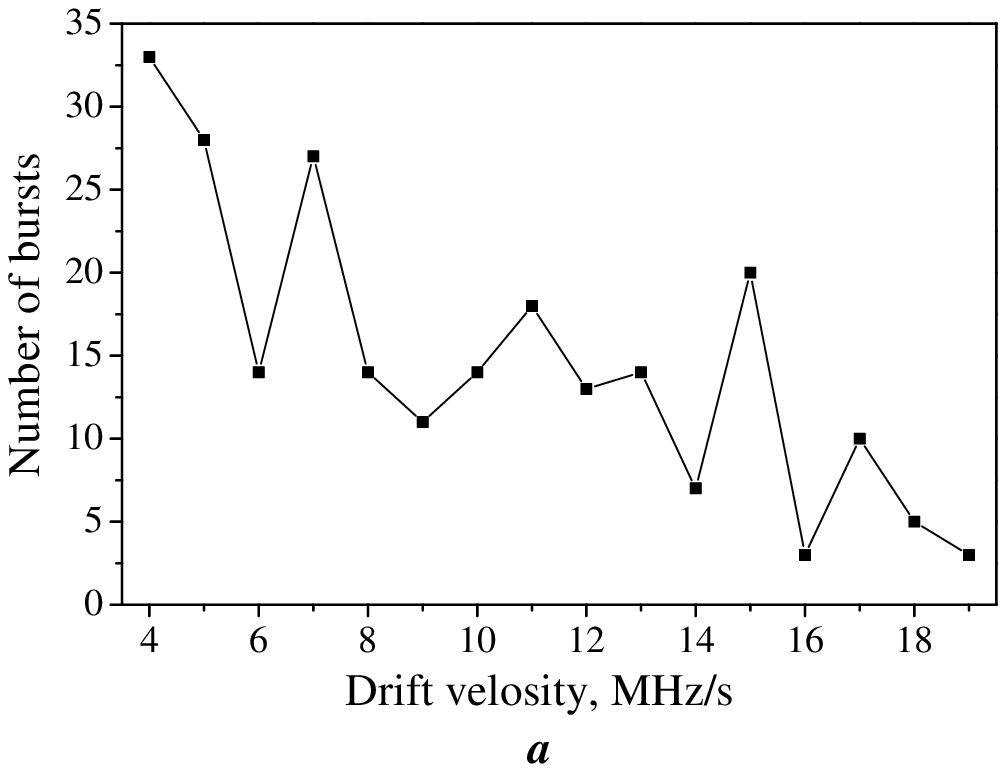} &
     \includegraphics[width=8cm]{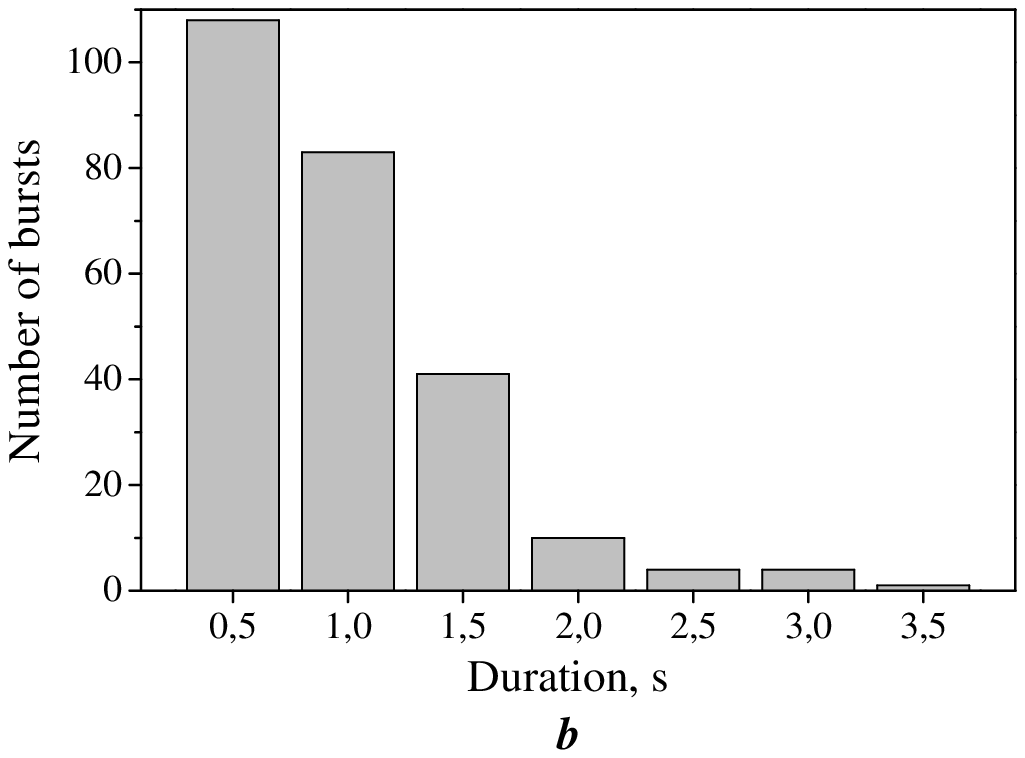}
\\  \end{array}
     $
     \end{center}
\caption{ (a)The frequency drift rate and, (b) duration distribution
of Type III-like bursts, observed in July 2003. \label{Fig3}}
\end{figure}

\begin{figure}
     \begin{center}
     $
     \begin{array}{cc}
     \includegraphics[width=8cm]{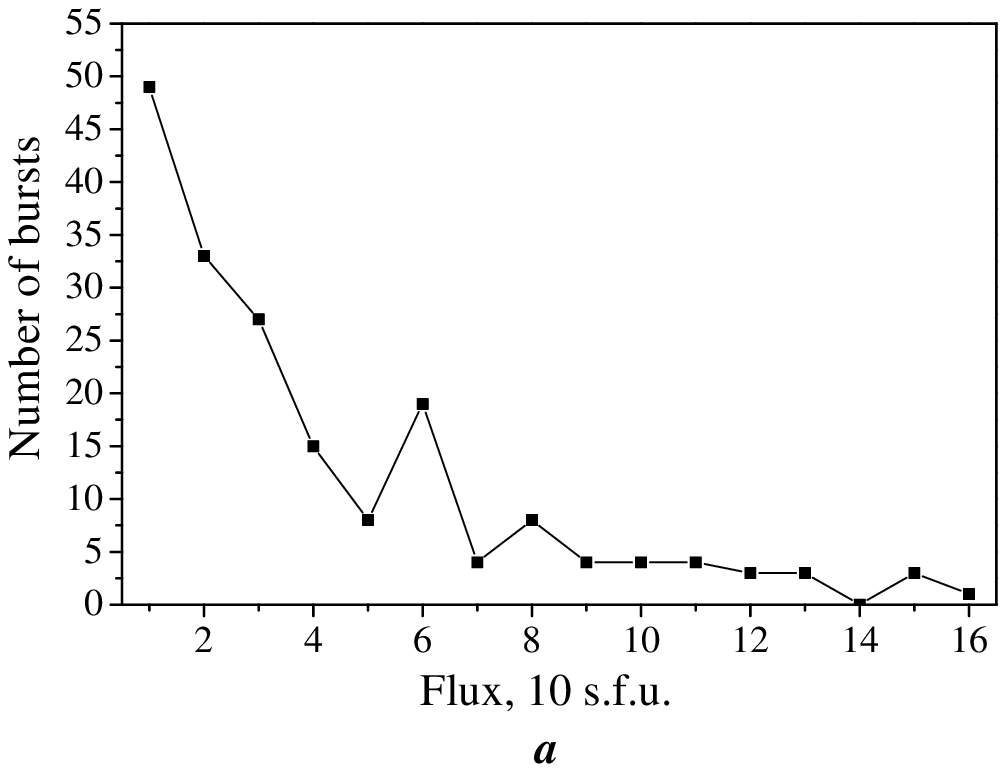} &
     \includegraphics[width=8cm]{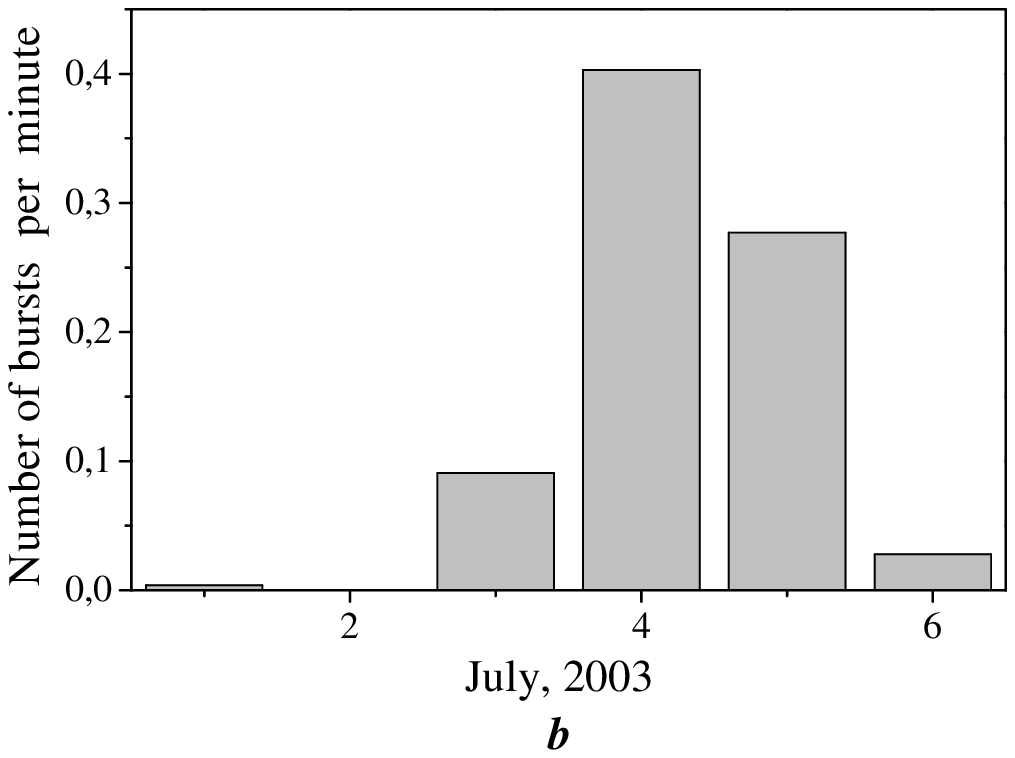}
\\
     \end{array}
     $
     \end{center}
\caption{ (a) The flux distribution of Type III-like bursts,
observed in July 2003, and (b) the appearance rate of Type III-like
bursts during five days from 1st to 6th of July 2003. \label{Fig4}}
\end{figure}

\section*{Analysis}
\indent \indent The big number of observed bursts allows statistical
analysis of Type III-like bursts. The distribution of Type III-like
bursts according to their frequency drift rates is shown in
Fig.~\ref{Fig3}a. The majority of bursts have frequency rates
12~MHz/s, though there are bursts with essentially larger rate
values.

One can see, that in the decameter range the drift rate of fast
bursts is few times bigger than that for usual bursts. A similar
relation between the fast and normal Type III bursts was observed at
higher frequencies \cite{Elgaroy80, Elgaroy74, Kundu61}.

The typical duration of decameter Type III-like burst is 1-2 s
(Fig.~\ref{Fig3}b), which is about 1/4 of the average duration of
ordinary decameter Type III bursts, their flux values are in the
range 10 to 60 s.f.u. (Fig.~\ref{Fig4}a). We analyzed 5 storms and
found that the majority of Type III-like bursts had been observed
when the active regions associated with them were located near the
central solar meridian, according to the data from Pulkovo
Observatory of Russian Academy of Sciences \cite{Pulkovo}. For
example in Fig.~\ref {Fig4}b the occurrence frequency of Type
III-like bursts within duration of one storm is shown. The maximum
on 4 July corresponds to the day, when the active region associated
with the Type III-like bursts was near to the central meridian. An
even more interesting case of two overlapping peaks on a similar
frequency occurrence diagram was observed when two active regions on
the Sun disk appear within a short time interval.

During the storm of June 17-22, 2004, the measurements of
polarizations were made for a part of Type III-like bursts by means
of radio telescope URAN-2 \cite{Brazhenko90}. The polarization
degree of Type III-like bursts proved to be equal to 10-15\%. There
were also observed Type III-like bursts with fine frequency
structure similar to usual Type IIIb bursts. Such Type III-like
bursts were not observed at higher frequencies, they are typical for
the decameter range only.

Up to now, all kinds of Type III and Type III-like bursts were
explained by the gradients of plasma density in the solar corona.
According to some models, the fast Type III-like bursts in the high
frequency range appear due to the fact that they are generated in
the low corona, where the density profile is steeper
\cite{Elgaroy80, Young61}. Following this model, the observed high
frequency drift rates of decameter Type III-like bursts forces us to
suppose that such density gradient is even higher in the corona, and
such irregularities must occupy very large regions (comparable with
the solar radius). However there is no evidence for existence of
such high density gradients in the corona. This leaves unanswered
the question about generation of big number of the Type III-like
bursts in the decameter range. The dependence of appearance rate of
Type III-like bursts via active region position suggests that the
propagation effects of electromagnetic waves in the solar corona
should be of big importance in this process.

\section*{Conclusions}
\indent \indent For the first time we discovered the Type III-like
bursts in decameter range. These bursts show properties similar to
those for Type III-like bursts observed at higher frequencies. For
example, their durations are 1/4 of usual decameter Type III bursts
durations, and their drift rates are few times of those for the
usual decameter Type III bursts. The number of Type III-like bursts
correlates with the position of the active region associated with
them. According to the polarization measurements of these bursts
(approximately 10-15\%), one can conclude that they seem to be
generated at the second harmonic of the plasma frequency. The
existing model of high frequency Type III-like bursts assume that
the plasma density gradients are much higher in the low corona. Such
assumption cannot be applied to the decameter bursts. The total
number of Type III-like bursts in the decameter range is remarkably
high (up to one bursts per 2.5 minutes), therefore they cannot be
explained by occasional extraordinary variations of solar
parameters, and there should exist a model based on typical values
of solar parameters.

Taking into account significant similarity between Type III-like bursts at
high frequency observed before and recently observed decameter Type III-like
burst (in particular the correlation between the position of the active
region on the solar disk and the burst occurrence rate) it may appear that
the suggested model of high frequency Type III-like burst should be
improved.

\end{document}